\documentclass[%
 reprint,
 superscriptaddress,
 showpacs,
 amsmath,amssymb,
 aps,
 pra,
 floatfix,
 longbibliography
]{revtex4-1}

\usepackage{graphicx}% Include figure files
\usepackage{dcolumn}% Align table columns on decimal point
\usepackage{bm}% bold math
\usepackage{braket}
\usepackage[dvipdfmx,colorlinks,linkcolor=blue,anchorcolor=blue,citecolor=blue, urlcolor=blue]{hyperref}

\begin{document}

%-------------------------------------------------------------------------------------------------

\title{Imaging the transverse spin density of light via electromagnetically induced transparency}
\author{Jinhong Liu}
\email[Corresponding author: ]{liujh\_jz@163.com}
\affiliation{Department of Science, Taiyuan Institute of Technology, Taiyuan 030008, China}
\author{Jinze Wu}
\affiliation{Interdisciplinary Center of Quantum Information, State Key Laboratory of Modern Optical Instrumentation, and Zhejiang Province Key Laboratory of Quantum Technology and Device of Physics Department, Zhejiang University, Hangzhou 310027, China}

%\date{\today}

%-------------------------------------------------------------------------------------------------

\begin{abstract}
When a light beam is strongly laterally confined, its field vector spins in a plane not perpendicular to the propagation direction, leading to the presence of transverse spin angular momentum, which plays a crucial role in the field of chiral quantum optics. The existing techniques to measure the transverse spin density require complex setups and sophisticated time-consuming procedures. Here, we propose a scheme to measure the transverse spin density of an optical field in real time using a multi-level atomic medium. The susceptibility of the medium is spatially modulated by the transverse spin via electromagnetically induced transparency. The distribution of the transverse spin is then extracted by measuring the distributions of the Stokes parameters of another collimated probe field.
\end{abstract}

%\pacs{42.50.Gy, 42.50.Ct}

\maketitle

%-------------------------------------------------------------------------------------------------

%\section{INTRODUCTION}

Light carries spin angular momentum, which is associated with circular polarizations. For a well-collimated light beam, which can be described within the paraxial approximation, the spin density is either parallel or antiparallel to the propagation direction of the beam. However, the spin density can be perpendicular to the propagation direction when the light experiences strong lateral confinement, which excites the longitudinal field component being $\pi/2$ out of phase with respect to the transverse component, implying a transversely spinning field vector \cite{Bliokh2015,Aiello2015}. Such transverse spin angular momenta are present in tightly focused beams \cite{Neugebaue2015,Neugebauer2018}, evanescent waves \cite{Bliokh2014}, multi-wave interference \cite{Bekshaev2015}, surface plasmon polaritons \cite{Bliokh2012}, whispering gallery modes \cite{Junge2013}, and near fields of nanostructures \cite{Saha2016}.

One of the most important properties of the transverse spin is the spin-momentum locking \cite{Mechelen2016}, which is a manifestation of quantum spin Hall effect of photons \cite{Bliokh2015-science}. When the light with transverse spin interacts with matters, the spin-momentum locking leads to a variety of propagation-direction-dependent emission and coupling effects, bringing about the field of chiral quantum optics \cite{Lodahl2017}. The transverse spin has been employed to implement single-atom controlled optical isolator \cite{Sayrin2015} and circulator \cite{Scheucher2016}, spin-dependent routing of single photons \cite{Shomroni2014}, nano photonic waveguide interface \cite{Petersen2014}. In addition, it has potential applications in optical tweezers \cite{PhysRevA.89.033841} and near-field microscopy \cite{Lee2006}.

Since the transverse spin density is a near-field three-dimensional polarization quantity, its measurement usually involves near-field and nanoprobing techniques \cite{Bauer2013}. One of the important methods to measure the transverse spin density employs a subwavelength nanoparticle on an interface as a local field probe. Scanning the nanoparticle and analyzing the scattered far field, one can reconstruct the distribution of the transverse spin density \cite{Neugebaue2015,Neugebauer2018}. However, such a method requires complex experimental setups and time-consuming measurement processes, hindering it widespread use in many applications requiring high real-time.

In this article, we propose a scheme to implement real-time imaging of transverse spin density of light using a multi-level atomic medium. The underlying mechanism is the so-called electromagnetically induced transparency (EIT) \cite{RevModPhys.77.633}, which is a quantum interference effect occurring in coherent systems \cite{PhysRevA.98.043829} and plays a crucial role in quantum information processing \cite{Monroe2002,Kimble2008}, precision metrology \cite{Hinkley2013,Budker2007}, preparation of quantum entanglement \cite{Horodecki2009}. The information of the transverse spin of a coupling field is mapped to the spatially-dependent susceptibility of the atomic medium, and then is extracted by another probe beam. The transverse spin density can be reconstructed from the distributions of the Stokes parameters of the transmitted probe beam. The proposed scheme is time-efficient and easy to be realized, and can be used in various applications related to transverse spin.

\begin{figure*}
\includegraphics[width=\textwidth]{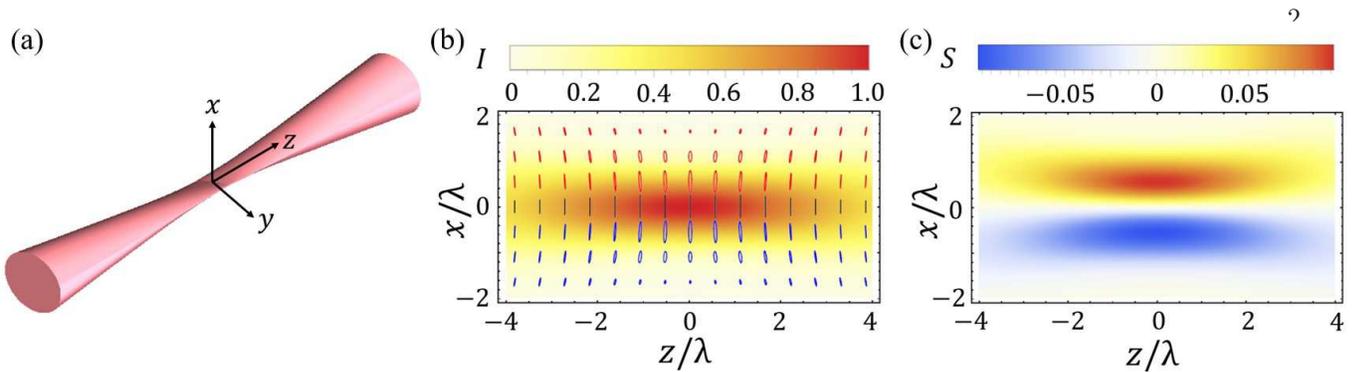}
\caption{
    (a) Gaussian beam tightly focused along the $x$-axis. (b) The distributions of intensity and polarization. The red and blue ellipses denote the right- and left-circular polarizations, while the black lines denote the linear polarization. (c) The distribution of the transverse spin density $\bm{\mathrm{s}}$ with $w_{0x}=10\lambda$ and $w_{0x}=10^4\lambda$.
}
\label{fig-01}
\end{figure*}

We consider a monochromatic Gaussian beam propagating along the $z$-axis and linearly polarized along the $x$-axis, as depicted in Fig.~\ref{fig-01}(a). It is tightly focused along the $x$-axis with a small beam waist of $w_{0x}$, while keeps well collimated along the $y$-axis with a much larger beam waist of $w_{0y}\gg w_{0x}$. Although both the electric and magnetic fields contribute to the spin angular momentum of the beam, we here only consider the electric field, because the electric dipole interaction dominates when the beam is coupled to atoms. The electric field distribution can be approximately expressed as:
\begin{equation}
    \bm{\mathrm{E}}(\bm{\mathrm{r}})={\cal E}_0\left(
    \hat{\mathrm{\mathbf{e}}}_x-\frac{x}{q_x}\hat{\mathrm{\mathbf{e}}}_z\right)u(\bm{\mathrm{r}}),
    \label{eq1}
\end{equation}
with
\begin{equation}
    u(\bm{\mathrm{r}})=\left(\frac{z_{Rx}z_{Ry}}{q_{x}q_{y}}\right)^{\frac{1}{2}}e^{ik\left(\frac{x^2}{2q_x}+\frac{y^2}{2q_y}\right)},
    \label{eq2}
\end{equation}
$q_x=z-iz_{Rx}$, and $q_y=z-iz_{Ry}$. Here ${\cal E}_0$ is the amplitude, $z_{Rx}=kw_{0x}^2/2$ and $z_{Ry}=kw_{0y}^2/2$ are the Rayleigh ranges. In order to satisfy the transverse constraint of Maxwell's equations, i.e., $\mathrm{\mathbf{\nabla}}\cdot \mathrm{\mathbf{E}}=0$, a longitudinal field component $E_z$ along the $z$-axis is induced due to the tight focusing. Note that there exists a nonzero phase difference between the transverse and longitudinal field components, resulting in a transversely spinning field vector in the $xz$-plane, as shown in Fig.~\ref{fig-01}(b). The field is right-, left-circularly, and linearly polarized in the regions of $x>0$, $x<0$, and $x=0$, respectively. This phenomenon is well described by the spin density of electric field defined as:
\begin{equation}
    \bm{\mathrm{s}}=\frac{\epsilon_0}{4\omega}\mathrm{Im}(\mathrm{\mathbf{E}}^\ast \times \mathrm{\mathbf{E}}),
    \label{eq3}
\end{equation}
with the vacuum permittivity $\epsilon_0$ and the angular frequency $\omega$ of the field. By substituting Eq.~(\ref{eq1}) into Eq.~(\ref{eq3}), we obtain
\begin{equation}
    \bm{\mathrm{s}}(\mathrm{\mathbf{r}})={\cal E}_0^2\frac{\epsilon_0}{4\omega}\frac{x}{q_x}|u(\mathrm{\mathbf{r}})|^2\hat{\mathrm{\mathbf{e}}}_y.
    \label{eq4}
\end{equation}
As we can see from Eq.~(\ref{eq4}), the spin density $\mathrm{\mathbf{s}}$ is purely transverse. Its longitudinal component vanishes due to the transverse linear polarization. It is parallel and antiparallel to the $y$-axis for $x>0$ and $x<0$, respectively, and becomes zero for $x=0$, corresponding to right-, left-circular, and linear polarizations, as illustrated in Fig.~\ref{fig-01}(c). It is to be noted that the large transverse spin density is present only in the region near the focal plane ($z=0$).

\begin{figure*}
\includegraphics[width=\textwidth]{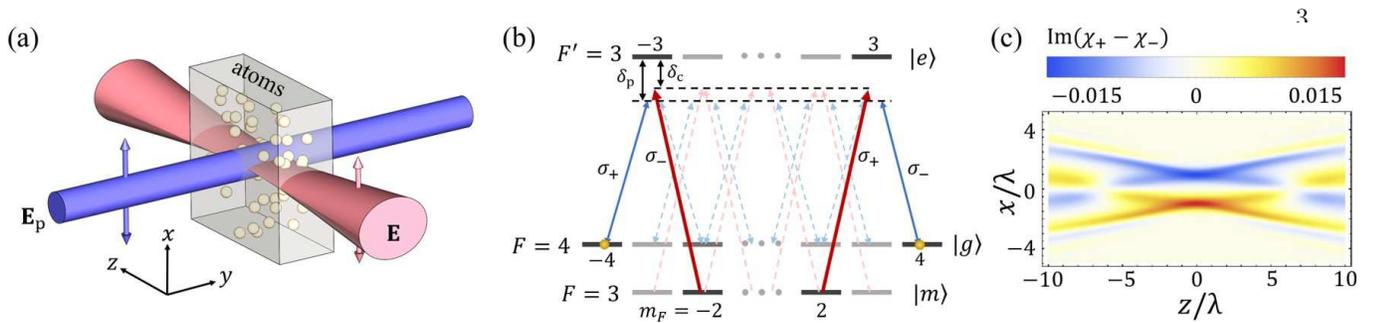}
\caption{
    (a) The transverse spin density $\mathrm{\mathbf{s}}$ of a tightly focused beam $\mathrm{\mathbf{E}}$ is encoded into the susceptibility of an $^{133}$Cs medium and then is extracted by a well-collimated probe beam $\mathrm{\mathbf{E_p}}$. (b) The relevant energy levels of $^{133}$Cs atoms prepared in the outer most Zeeman sublevels $\ket{g,m_F=\pm 4}$. (c) The distribution of $\mathrm{Im}(\chi_{+}-\chi_{-})$ characterizing the absorption difference between the two opposite spin components of the probe beam $\mathrm{\mathbf{E_p}}$. The parameters are $N=4.2\times 10^{25}\,\mathrm{m}^{-3}$, $\Omega=3\Gamma$, $\delta_c=\delta_p=0$, $\gamma_{eg}=0.6\Gamma$, and $\gamma_{mg}=0.1\Gamma$. Here $\Gamma$ is the spontaneous decay rate.
}
\label{fig-02}
\end{figure*}

When this beam is coupled to an atomic medium, its transverse spin distribution is encoded into the atomic susceptibility [see Fig.~\ref{fig-02}]. To reveal the underlying physics, we consider an $^{133}$Cs medium with two ground states $\ket{g}=\ket{6\mathrm{S}_{1/2},F=4}$ and $\ket{m}=\ket{6\mathrm{S}_{1/2},F=3}$, and an excited state $\ket{e}=\ket{6\mathrm{P}_{1/2},F'=3}$ in the D1 line, as shown in Fig.~\ref{fig-02}(b). The atoms are assumed to be prepared in the outer most Zeeman sublevels $\ket{g, m_F=\pm 4}$ by a resonant incoherent pump fields (not shown in Fig.~\ref{fig-02}) linearly polarized along the $y$-axis, which is taken to be the quantization axis. The coupling field can be decomposed into two spin components:
\begin{align}
    \bm{\mathrm{E}}(\mathrm{\mathbf{r}})=&\bm{\mathrm{E}}_{+}(\mathrm{\mathbf{r}})+\bm{\mathrm{E}}_{-}(\mathrm{\mathbf{r}})\nonumber\\
    =&\frac{{\cal E}_0}{\sqrt{2}}\left(1+\frac{ix}{q}\right)u(\mathrm{\mathbf{r}})\hat{\mathrm{\mathbf{e}}}_{+}+\frac{{\cal E}_0}{\sqrt{2}}\left(1-\frac{ix}{q}\right)u(\mathrm{\mathbf{r}})\hat{\mathrm{\mathbf{e}}}_{-},
    \label{eq5}
\end{align}
with $\hat{\mathrm{\mathbf{e}}}_{\pm}=\frac{1}{\sqrt{2}}(\hat{\mathrm{\mathbf{e}}}_{x}\pm i\hat{\mathrm{\mathbf{e}}}_{z})$. The two components drive the $\sigma^+$- and $\sigma^-$-transitions between $\ket{e}$ and $\ket{m}$,  respectively. A well-collimated probe beam propagating along the $y$-axis and linearly polarized along the $x$-axis is also coupled to the atomic medium. Since all the atoms are in the sublevels $\ket{g, m_F=\pm 4}$, the two spin components of the probe beam can only drive the $\sigma^+$-transition $\ket{e, m_{F^{'}}=-3}\leftrightarrow \ket{g, m_F=-4}$ and the $\sigma^-$-transition $\ket{e, m_{F^{'}}=3}\leftrightarrow \ket{g, m_F=4}$, and experience susceptibilities expressed as:
\begin{equation}
   \chi_+=\frac{7Nd_{eg}^2}{12\epsilon_0\hbar}\frac{(\delta_p-\delta_c-i\gamma_{mg})\rho_{g_{-4,-4}}}{(\delta_p-\delta_c-i\gamma_{mg})(\delta_p-i\gamma_{eg})-|\Omega_-(\mathrm{\mathbf{r}})|^2},
    \label{eq6}
\end{equation}
\begin{equation}
   \chi_-=\frac{7Nd_{eg}^2}{12\epsilon_0\hbar}\frac{(\delta_p-\delta_c-i\gamma_{mg})\rho_{g_{4,4}}}{(\delta_p-\delta_c-i\gamma_{mg} )(\delta_p-i\gamma_{eg})-|\Omega_+(\mathrm{\mathbf{r}})|^2},
    \label{eq7}
\end{equation}
with the atomic number density $N$, the reduced dipole matrix element $d_{eg}$, the reduced Planck constant $\hbar$, the frequency detuning $\delta_c(\delta_p)$ of the coupling (probe) beam, the decoherence rate $\gamma_{em}(\gamma_{eg})$ of the transition $\ket{e}\leftrightarrow\ket{m}(\ket{e}\leftrightarrow\ket{g})$, the population $\rho_{g_{-4,-4}}(\rho_{g_{4,4}})$ of the sublevel $\ket{g,m_F=-4}(\ket{g,m_F=4})$, and the Rabi frequency
\begin{equation}
   \Omega_{\pm}(\mathrm{\mathbf{r}})=\frac{\mathrm{\mathbf{d}}_{\pm} ^{\ast} \cdot \mathrm{\mathbf{E}}_{\pm}(\mathrm{\mathbf{r}})}{\hbar}
   =\frac{1}{4\sqrt{2}}\left(1 \pm \frac{ix}{q}\right)u(\mathrm{\mathbf{r}})\Omega,
    \label{eq8}
\end{equation}
where $\Omega=d_{em}{\cal E}_0/\hbar$ is the reduced Rabi frequency.
\begin{figure}
\includegraphics[width=\linewidth]{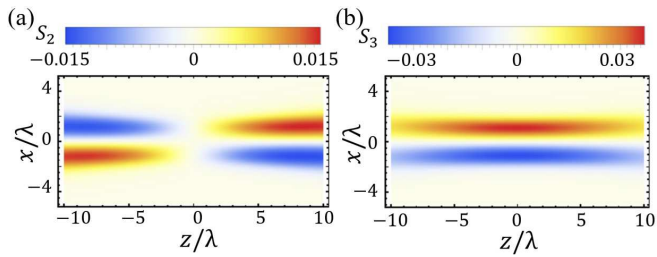}
\caption{
    The distributions of (a) $S_2$ and (b) $S_3$ of the transmitted probe beam. The parameters are the same in Fig.~\ref{fig-02}.
}
\label{fig-03}
\end{figure}

It can be seen from the above equations that the information of the polarization and thus the transverse spin of the coupling beam is encoded into the spatially-dependent susceptibility $\chi_{\pm}(\mathrm{\mathbf{r}})$ via the Rabi frequency $\Omega_{\pm}(\mathrm{\mathbf{r}})$. Due to the transverse spin, the atomic medium exhibits different susceptibilities for the two spin components of the probe beam, i.e., $\chi_+\neq \chi_-$,  as shown in Fig.~\ref{fig-02}(c). After passing through the medium, the probe beam carries a spatially-dependent polarization, which is characterized by the Stokes parameters. Fig.~\ref{fig-03} displays the distributions of the second and third Stokes parameters $S_2$ and $S_3$. In comparison with Fig.~\ref{fig-01}(c), a direct correspondence between the transverse spin density $\mathrm{\mathbf{s}}$ and the Stokes parameter $S_3$ can be found. Therefore, one is able to detect the distribution of $\mathrm{\mathbf{s}}$ of the coupling beam via mearing the $S_3$ distribution of the probe beam. Since the atomic medium has very fast response to the polarization change of light, this method can realize the real-time detection of the transverse spin density. In addition, the second Stokes parameter $S_2$ gives the orientation of the polarization ellipse [comparing Fig.~\ref{fig-01}(b) and Fig.~\ref{fig-03}(b)].

In conclusion, we have proposed an atom-based scheme to measure the transverse spin density of light. We have shown that the light with transverse spin can dramatically modify the susceptibility of an atomic medium. When a well-collimated probe beam passes through the medium, its polarization distribution carries the information of the transverse spin. One can measure the distributions of its Stokes parameters to obtain the distribution of the transverse spin density in real time.

%>>>>>>>>>>>>>>>>>>>>>>>>>>>>>>>>>>>>>>>>>>>>>>>>>>

%-------------------------------------------------------------------------------------------------

\bibliography{references}

\end{document}